\documentclass{vgtc}                          

\graphicspath{{figures/}{pictures/}{images/}{./}} 

\usepackage{times}                     

\usepackage{tabu}                      
\usepackage{booktabs}                  
\usepackage{lipsum}                    
\usepackage{mwe}                       
\usepackage{enumitem}

\usepackage{mathptmx}                  

\onlineid{1105}

\vgtccategory{Research Poster}

\vgtcinsertpkg

\title{Efficiently Crowdsourcing Visual Importance with Punch-Hole Annotation}

\author{
    Minsuk Chang$\thanks{e-mail: \{minsuk, cxiong\}@gatech.edu}${ }$^1$, %
    Soohyun Lee$\thanks{e-mail: \{shlee, archo, hj, shpark, jseo\}@hcil.snu.ac.kr}${ }$^2$,
    Aeri Cho$^\dag${ }$^2$,
    Hyeon Jeon$^\dag${ }$^2$,
    Seokhyeon Park$^\dag${ }$^2$, \\
    Cindy Xiong Bearfield$^*${ }$^1$,
    Jinwook Seo$^\dag${ }$^2$
}
\affiliation{\scriptsize $^1$Georgia Institute of Technology \quad $^2$Seoul National University}
     
\teaser{
  \centering
    \includegraphics[width=\linewidth]{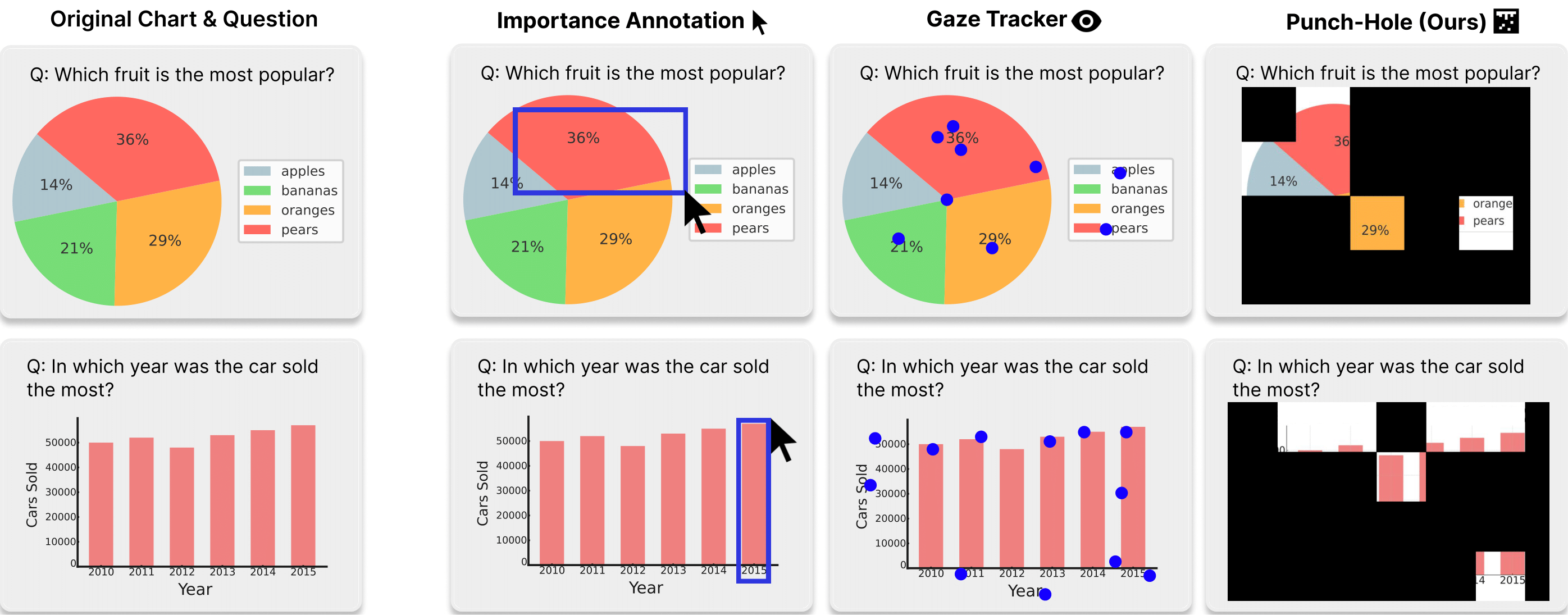}
    \caption{Comparison of three methods used to extract important areas in charts. \textbf{Importance Annotation}~\cite{turkeyes} (left): Drawing a box or polygon with a mouse. \textbf{Gaze Tracker}~\cite{scannerDeeply} (middle): Capture the points where the eye stays long. \textbf{Punch-Hole Labeling} (right): Extract important patches for proper question answering. The example annotation for Importance Annotation and Punch-Hole labeling is retrieved from the pilot study.}
  \label{fig:teaser}
}

\abstract{
    We introduce a novel crowdsourcing method for identifying important areas in graphical images through punch-hole labeling. Traditional methods, such as gaze trackers and mouse-based annotations, which generate continuous data, can be impractical in crowdsourcing scenarios. They require many participants, and the outcome data can be noisy. In contrast, our method first segments the graphical image with a grid and drops a portion of the patches (punch holes). Then, we iteratively ask the labeler to validate each annotation with holes, narrowing down the annotation only having the most important area. This approach aims to reduce annotation noise in crowdsourcing by standardizing the annotations while enhancing labeling efficiency and reliability. Preliminary findings from fundamental charts demonstrate that punch-hole labeling can effectively pinpoint critical regions. This also highlights its potential for broader application in visualization research, particularly in studying large-scale users’ graphical perception. Our future work aims to enhance the algorithm to achieve faster labeling speed and prove its utility through large-scale experiments.
} 


\begin{document}


\firstsection{Introduction}
\maketitle

Extracting saliency in graphical images has gained importance for analyzing cognitive activities, such as perceptual bias~\cite{sameData} and subjective importance~\cite{importanceLabeling}. Various researchers use crowdsourcing to retrieve and utilize such human-centric data on a large scale~\cite{sameData,scannerDeeply}.

However, crowd workers often annotate images directly (e.g., bounding boxes, polygons) or indirectly (e.g., gaze capture). While intuitive, this method introduces two main problems. First, labelers may overemphasize certain regions, neglecting other important areas. For example, TurkEyes~\cite{turkeyes} shows that the extracted importance in a resume is concentrated near the title, overshadowing other sections. Second, free-form annotations (e.g., colored pixels, gaze points) are highly variable, requiring data processing or combining responses from multiple participants. ScannerDeeply~\cite{scannerDeeply} addressed this by implementing active noise removal in gaze data to counter low accuracy of webcam-based gaze trackers, adding to processing efforts and raising labeling costs.

As a breakthrough, we suggest a novel grid-based annotation to effectively crowdsource the importance or saliency of the graphical image. Punch-hole labeling reduces the continuous annotation task into multiple binary questions, fostering consensus among participants and reducing labeling costs. The process involves dividing the image into small grids and hiding one patch at a time (similar to punching holes) to identify less important areas. If users can answer the given question based on the shown patches, it indirectly indicates that the punched holes lack significance. This approach aims to prevent omitting critical but overlooked regions, minimize unnecessary discrepancies among annotations, and simplify the annotation task for greater time efficiency.

Our preliminary study with two fundamental charts in \autoref{fig:teaser} suggests that Punch-hole labeling may achieve faster labeling speed and produce a more reliable importance map than traditional approaches. Our next goal is to enhance the punching algorithm to increase labeling speed and deploy a large-scale experiment in the crowdsourcing platforms.

\section{Punch-Hole Labeling} \label{sec:patchLabeling}
Punch-Hole annotation is based on the idea that \textbf{simplified questions and responses} in a crowdsourcing environment can improve both the speed and reliability of labeling tasks~\cite{simplerBetter}. Compared to the original annotation task of ``\textit{Annotate the important area related to this question.}'', it is reduced to the binary question of ``\textit{Can you answer the question based on these patches?}''. Our punch-hole algorithm and its advantages follow this principle.

\subsection{Algorithm} \label{sec:Algorithm}
We begin by dividing an image into square patches using a predefined grid size and displaying only a subset of these patches to crowd workers. This approach involves sequentially hiding each patch, like punching holes in a piece of paper, and asking whether the remaining visible portions of the image are sufficient to answer the given question. In \autoref{fig:teaser}, the remaining areas are essential for answering the questions, compared to unnecessary black patches. For instance, the pie chart's top-right corner or the bar chart's middle, which appears black, is unimportant for retrieving the answer. 

This process is iterative, with holes being punched until all visible patches are crucial for the given question. Subsequently, the area’s resolution can be enhanced by reducing the grid size and repeating the process, theoretically achieving pixel-level granularity. Labeling time and patch size present an inverse tradeoff that deployers can adjust based on their budget or quality requirements.

\subsection{Advantages}
The potential advantages of punch-hole labeling can be expressed with three key terms:
\begin{itemize}[left=0pt, itemsep=0pt]
    \item \textbf{Show-And-Verify} strategy prevents underestimation of specific areas. Punch-hole labeling controls the information presented to users, ensuring no sections are overlooked. Previous methods using active responses like gaze points and polygons~\cite{scannerDeeply, turkeyes} allowed too much freedom, leading to incomplete coverage. For example, users might skim the legend and focus solely on the data, neglecting the legend’s importance. Our approach includes all crucial patches in the final annotation, enhancing quality.

    \item \textbf{Discrete Responses} reduce the number of required participants. Previous methods like gaze trackers~\cite{scannerDeeply} and mouse-based interfaces~\cite{turkeyes} face issues with merging diverse responses, requiring many participants to suppress noise~\cite{turkeyes}. Punch-hole annotations, with lower granularity and standardized format, separate controversial or subjective areas from consensus ones. This method achieves the desired granularity after iterations with smaller patch sizes, reducing the impact of outliers and noise.

    \item \textbf{Task Simplicity} increases accessibility. Punch-hole labeling simplifies and standardizes the task, requiring only two buttons (yes/no). Traditional tools require facial position calibration~\cite{scannerDeeply} or precise clicking and brushing~\cite{turkeyes}, relying heavily on motor skills. This can limit accessibility for elderly users, significant contributors to crowdsourcing. Poor devices can also produce noisy annotations, requiring iterations for accuracy~\cite{scannerDeeply}. The punch-hole algorithm addresses this by focusing on the cognitive task, reducing complexity, and enhancing accessibility.

\end{itemize}
\section{Preliminary Study}
As a preliminary study, we tested the effectiveness and efficiency of our punch-hole annotation with a pilot test with two labelers. We analyzed whether our approach could effectively find the important area in the image while reaching a faster labeling speed.
\subsection{Procedure}
We started our study by generating the fundamental graphical image for the experiment. Leveraging ChatGPT, we created one pie chart with fruit sales and one bar chart with car sales. Trailing questions were also generated based on the charts' contents. We then tested our algorithm on a simple web-based labeling tool and compared its results with the box annotation from importance annotation~\cite{turkeyes}. Users were allowed to annotate more than one box for a fair comparison. The time spent to label each annotation was recorded for later analysis.

\subsection{Results}
\begin{itemize}[left=0pt, itemsep=0.5pt]
    \item \textbf{Extracted Annotations.} An example of punch-hole annotation is shown in \autoref{fig:teaser}. Punch-hole annotations include the core area needed to answer the question correctly. In importance annotation, users’ marked areas varied greatly; one marked the whole bar, another only the top. Users also often miss or over-mark legends as important.

    \item \textbf{Time Analysis.} The average labeling speed was 1.32 seconds per annotation candidate and 30.36 seconds per chart. Punch-hole labeling takes a similar time to the 30 seconds reported for importance annotation~\cite{turkeyes}. By dynamically adjusting the punch-hole shape and refining which patches to hide, we aim to create more precise areas in less time.
\end{itemize}

\subsection{Discussion \& Future Work}
We have confirmed that punch-hole labeling can extract relative importance from the image within a reasonable time. Our findings also indicate that traditional annotation techniques often miss essential areas and produce highly variable results. Consequently, we aim to advance our punch hole algorithm by: 1) minimizing the number of punch holes to reduce the number of micro-tasks for crowdsourcing workers, and 2) dynamically determining the optimal punch-hole size for each image.

We also observed that the order in which we punch holes may influence the final output. Randomizing the order of punching and analyzing how the final areas are extracted will be another future direction. To achieve this, we plan to experiment with this approach in a crowdsourcing platform to increase the reliability of the results and conduct a large-scale analysis of the extracted areas, ultimately forming a dataset. We expect that our work will offer researchers focused on human cognition and visual perception the opportunity to crowdsource large-scale annotations, even with limited budgets.

\acknowledgments{
This work is supported by NSF awards IIS-2237585 and IIS-2311575.
}

\bibliographystyle{abbrv-doi}

\bibliography{template}
\end{document}